# Influence of surface structure and strain on indium segregation in InGaAs: a first-principles investigation

Soumya S. Bhat[1], Satadeep Bhattacharjee[1], Jung-Hae Choi[2], Seung-Cheol Lee[1*]

[1]Indo-Korea Science and Technology Center (IKST), Bangalore 560065, India.

[2]Center for Electronic Materials, Korea Institute of Science and Technology, Seoul 02792, Republic of Korea.

**Abstract**

The surface segregation of indium atoms in InGaAs is investigated using first-principles calculations based on density functional theory. Through the calculation of segregation energies for (100), (110), and (111) surfaces of GaAs we analyze the decisive role of surface orientation on indium segregation. Further, our calculations reveal that the variation of segregation energy as a function of applied strain strongly depends on surface reconstruction. Obtained segregation energy trends are discussed in light of atomic bonding probed via integrated crystal orbital Hamilton population. Results presented in this paper are anticipated to guide experimental efforts to achieve control on In segregation by managing As-flux along with the application of strain.

## 1. Introduction

III-V semiconductor ternary alloys have received extensive attention due to their technological significance in the semiconductor industry. They are increasingly used in optoelectronic devices as they offer specific bandgaps by adjusting the relative compositions of the constituent elements to achieve high performance. Despite recent advances in crystal growth techniques like molecular beam epitaxy (MBE) to obtain high-quality III-V semiconductors, physical limitations such as growth islands, interface diffusion, and surface segregation are inevitable [1]. The occurrence of surface segregation during the growth of III-V ternary semiconductor alloys and heterojunctions between two given binary III-V compounds has been demonstrated by various surface sensitive techniques such as Auger electron spectroscopy (AES), x-ray photoemission spectroscopy (XPS) and reflection high-energy electron diffraction (RHEED) [2–4]. Preferential segregation of one of the third column atoms at the surface, to the expense of the other, within their common sublattice leads to bulk-surface redistribution. As a result, the surface is found to be nearly binary in ternary alloys, and interface roughening occurs in heterojunctions [1]. In the particular case of InGaAs, during MBE growth, experiments revealed preferential segregation of indium, leading to an excess of indium at the surface [5–8].

Segregation inherently restricts the formation of abrupt interfaces necessary for the production of high-performance devices. To improve interface abruptness, one need to kinetically hinder the In surface segregation during the growth of InGaAs. This has been achieved by controlling growth conditions, as In segregation is expected to be strongly dependent on the growth rate, temperature, V/III equivalent pressure ratio, alloy compositions, and /or strain. Many experimental studies demonstrated that the best way to limit segregation is to prevent the system from reaching thermodynamic equilibrium by using low growth temperatures or high growth rates [2,9,10]. Further In segregation is also found to be largely influenced by surface reconstruction [11]. For instance, it is well established that In segregation can be suppressed by increasing the As to group III pressure ratio, since In segregation is reduced for an As-stabilized reconstructed surface as compared to group III stabilized surface [9,11–13].

As a result of large lattice constant mismatch (~7%) between InAs and GaAs, InGaAs/GaAs heterostructures are strained unless the stress caused by the pseudomorphic deformation is relaxed by the formation of defects such as dislocations [14]. Elastic strain as a consequence of lattice constant mismatch could be one of the important aspect affecting the segregation. However, conflicting results are reported by experiments concerning the effect of strain on In segregation. Earlier investigations by Moison et al. [15] on MBE grown heterostructures showed a marginal change in In segregation for a compressive strain ranging from 0 to 7%. This was supported by Grenet et al. [8], who performed XPS investigation of In surface segregation for As-stabilized GaInAs ternary bulk alloys grown by MBE. Their experiments revealed no direct effect of strain on In segregation for $Ga_{0.25}In_{0.75}As$. While, contradictory results were reported by Garcia et al. [16], where authors examined the strain relaxation and In segregation effects during MBE growth of InAs on GaAs(001), and proposed a more complex growth mode in which stress energy induces In segregation. Later many experiments disclosed the significant influence of strain on In segregation [17–19]. This discrepancy in the experimental findings regarding the effect of strain on In segregation makes it imperative to carry out a thorough analysis.

To elucidate the role of strain on In segregation, we study the effect of strain on In segregation using first-principles calculations based on density functional theory (DFT).

Objectives of this work are two-fold: i) to investigate the influence of surface reconstruction on In segregation and, ii) effect of strain on segregation behavior, so as to account for discrepancies in the earlier experimental results. Through these calculations, the influence of group V/III flux and strain on In segregation has been established. We provide a quantitative comparison of surface segregation energies for various surface orientations and under applied bi-axial strain with implications to observed experimental trends. Further, obtained results are described by atomic bonding analysis.

## 2. Computational details

We performed first-principles calculations using the projector-augmented wave (PAW) [20] method in the framework of DFT as implemented in the Vienna Ab initio Simulation Package (VASP) code [21,22]. The exchange-correlation energy of electrons is treated within a generalized gradient approximated functional (GGA) of the Perdew-Burke-Ernzerhof (PBE) [23] parametrized form. Interactions between ionic cores and valence electrons are represented using PAW pseudopotentials, where 3d (4d), 4s (5s), 4p (5p) electrons for Ga (In) and 4s, 4p electrons for As are treated as valence. Plane-wave basis set with a kinetic energy cutoff of 500 eV and an energy convergence criteria of $10^{-6}$ eV are used. The optimized lattice constants for bulk GaAs and InAs are 5.75 and 6.19 Å respectively, which are in reasonable agreement with the reported experimental values of 5.65 and 6.06 Å, respectively [24].

The four most stable surfaces of GaAs are modeled by slabs in the periodic supercells, with a vacuum thickness of ~ 15 Å for separating the slabs. The slabs consist of nine GaAs layers. The atoms in the bottom two layers are fixed at the bulk lattice sites, while all other atoms are allowed to relax until the forces on those atoms are less than 0.02 eV/Å. Gamma-centered k-point grid was used for Brillouin zone sampling. Dipole correction is applied in order to avoid artificial interactions between periodic slab images. The In atoms are substituted in the Ga positions of upper five layers and all possible single substitutions are investigated.

To understand the segregation behavior of In in GaAs, the segregation energy is determined as the difference in the total energies of the structures with the In atom located at the surface atomic layers and in the bulk. Accordingly, the surface segregation energy $E_{segr}$ is estimated by the following equation,

$$E_{segr} = E_{InGaAs(In, nth\ layer)} - E_{InGaAs(In, 5th\ layer)} \qquad (1),$$

where $E_{InGaAs\ (In,nth\ layer)}$ and $E_{InGaAs\ (In,5th\ layer)}$ denotes the total energies of the InGaAs alloy system with the In atom located at the n$^{th}$ Ga layer (n = 1 - 4) and 5$^{th}$ Ga layer respectively. For 5$^{th}$ Ga layer, we assume that the In atom is in the 'bulk' like environment as placing In in lower Ga layers has a marginal effect on the segregation energy. A positive value of $E_{segr}$ implies that the In atom tends to sit inside the GaAs bulk, while a negative value indicates that the In atom segregates towards the surface.

## 3. Results and discussion

### 3.1 Segregation energy

*3.1.1 The (110) surface*

The (110) cleavage surface is one of the most both experimentally and theoretically studied GaAs surface. It is the ideal surface for theoretical investigations as it does not reconstruct, but retains its primitive (1×1) symmetry. The surface relaxes with As atoms moving outwards from their ideal positions and Ga atoms moving inward, with a tilt angle of ~30° [25,26].

It is well established that surface energy difference plays a crucial role in determining the segregation behavior in binary alloys [27,28]. So, as a first step we have determined the surface energies for (110) surfaces of GaAs and InAs using both GGA and LDA. As (110) cleavage surface is stoichiometric, its surface energy is independent of chemical potentials. The specific surface energy γ is calculated using the following equation [29,30],

$$\gamma = \frac{1}{2}[E_{slab}(N) - NE_{bulk}] \qquad (2)$$

where '2' in the denominator included as two surfaces are involved in the calculation, $E_{slab}$ and $E_{bulk}$ are the total energy of the slab with *N* number of layers and that of bulk, respectively.

For GaAs(110), calculated surface energies are 0.57 and 0.42 eV/atom using LDA and GGA respectively. When expressed in terms of meV/Å$^2$, i.e. surface free energy per unit area, it is 51 and 36 meV/Å$^2$, respectively. This is in reasonable agreement with the experimental value of 0.61 eV/atom (54 meV/Å$^2$) determined from the fracture experiments [24] and with the earlier theoretical calculations [30,31]. For InAs(110) we obtain the surface energy of 0.52 eV/atom (40 meV/Å$^2$) and 0.38 eV/atom (28 meV/Å$^2$) within LDA and GGA respectively. This agrees with the previous DFT results by Choudhury et al. [31]. Note that LDA works better than

GGA for surface energies which have been confirmed earlier [30] due to better cancellation of errors between surface exchange and correlation energies [32].

Above results demonstrate that it is energetically favorable for In atom to segregate towards the GaAs(110) surface as its surface energy is lower than that of InAs(110) [27,33]. Figure 1 shows the schematic of In atom substituting for a Ga atom in the top-most and fifth atomic layers. The segregation energy for In atom in GaAs(110) and for Ga atom in InAs(110) are calculated using equation (1) and are illustrated in figure 2. As anticipated, for the case of In in GaAs(110) $E_{segr}$ is negative, indicating the tendency of In segregation to the surface, while for Ga in InAs(110) $E_{segr}$ is positive, implying Ga prefers to stay inside the bulk.

*3.1.2 The (100) surface*

Among all the surface orientations, the (100) surface of GaAs is of most technological interest as they are commonly used in the MBE growth method. It consists of alternating Ga and As planes separated by 1.41 Å [34], with the possibility of both As and Ga-terminations depending on the growth conditions. Apart from its application aspect, extensive theoretical and experimental efforts are devoted to establish its structural stability as it exhibits a large number of surface reconstructions depending on the surface preparation conditions [25,34–39]. It is now well established that for growth conditions ranging from As-rich to Ga-rich, the stable reconstructions of GaAs(100) surfaces follow the sequence c(4×4), $\beta$2(2×4), $\alpha$2(2×4), and $\zeta$(4×2) [37–39]. In the present work we considered c(4×4) and $\beta$2(2×4) structures which are identified as most stable arrangements under increasing As-rich conditions [40].

The c(4×4) reconstruction is formed by adding three As ad-dimers on top of a complete As monolayer. As the As-chemical potential decreases, the surface transforms into the $\beta$2(2×4) structure which has two As dimers in the topmost layer and a third As dimer in the third layer. In the above models, as the surface is terminated with As atoms, dangling bonds are easy to buckle together to form the As dimers [37,41]. The Ga-terminated bottom surface of the slab is saturated with fractionally charged pseudo-hydrogen with an atomic number of Z = 1.25, to mimic bulk As bonding.

### 3.1.3 The (111) surface

The ideal bulk truncated GaAs(111) is Ga terminated and undergoes surface reconstructions depending on the growth environment. Experiments revealed a (2×2) reconstruction, which is attributed to a Ga vacancy. The Ga vacancy model (111)*A*-Ga vacancy is the most favorable reconstruction for a wide range of the chemical potential from a Ga-rich to an As-rich environment [42]. An As-trimer model (111)*B*(2×2) is reported by experiments and theoretical studies only under extreme As-rich environments [25,43]. In this work we modeled GaAs(111)*A*-Ga vacancy structure, with As-terminated bottom surface passivated by pseudo-hydrogen with an atomic number of Z = 0.75, to imitate bulk Ga atom.

The segregation energy of single In atom in four different surface structures of GaAs determined using equation (1) is as shown in figure 3. For all surfaces considered, $E_{segr}$ is negative, confirming the tendency of In segregation towards the surface for all the four structures. Our results disclose the strong dependence of In surface segregation energies on the surface orientation of GaAs, where there are differences of the order of 0.1 eV for one In atom substitution. A similar dependency on the surface orientation is reported for transition metal alloys [27]. Calculated segregation energies can be used to estimate the monolayer coverages of the surfaces at finite temperatures using Langmuir-McLean theory of segregation which approximates segregation by balancing adsorption and desorption rates [44]. With the assumption that the surface segregation energy is independent of solute concentration for dilute concentration limit and the segregation is limited to one monolayer, Langmuir-McLean isotherm reads,

$$\frac{C_s}{1-C_s} = \frac{C_b}{1-C_b} \exp(-\frac{E_{segr}}{kT}) \qquad (3)$$

where $C_s$ is the occupancy of surface sites by solute atoms, $C_b$ is the site occupancy by solute atoms in the bulk at temperature $T$ and $k$ is the Boltzmann constant. The surface coverage reduces with temperature and the rate of reduction exponentially depend on the segregation energies [45]. The estimated surface coverage of indium at 300 K for four surfaces of GaAs are shown in figure 4. As seen from the figure, segregation of indium to the surface forming a complete monolayer of InAs is achieved at lower $C_b$ for (100)c(4×4) and (111)Ga-vacancy surfaces than for (110) cleavage and (100)$\beta$2(2×4). This implies, In surface segregation can be hindered by using (110) or (100)$\beta$2(2×4)) surfaces of GaAs so as to avoid abrupt interface formation at low indium concentrations.

For the specific case of (100) surface, the results presented in the previous section indicate that the tendency of In segregation towards the surface is more for c(4×4) reconstruction than for $β2(2×4)$, which contradicts the earlier experimental results. In 1992, Muraki et al. [9] investigated the surface segregation of In atoms during MBE growth of InGaAs/GaAs quantum wells (QWs) and showed that In surface segregation is less pronounced for high As-pressure than for the low-As pressure. This was the first work that demonstrated the strong dependency of As-overpressure on In segregation, which was attributed to the enhanced incorporation rate of In atoms under an increased surface coverage of As atoms. That is, the mobile In atoms on the growing surface have a shorter lifetime before they are incorporated into the bulk phase, making In surface segregation less pronounced. This was supported by the work of Nagle et al. [46] where they studied surface segregation of In in strained $Ga_{0.8}In_{0.2}As$/GaAs QWs by means of RHEED combined with ultraviolet photoelectron spectroscopy (UPS) measurements and revealed high sensitivity of segregation energy to the As flux than to the growth temperature. Similar results were observed by Ekenstedt et al. [11] in their RHEED measurements on strained $In_{0.5}Ga_{0.5}As$ layers grown on InAs. Their experiments disclosed an effective reduction in In segregation for an As-covered (2×4) reconstructed surface as compared to a group III stabilized (4×2) surface reconstruction.

A plausible reason for this discrepancy between earlier experiments and our calculations could be the effect of strain, which has not been considered so far in our calculations.

## 3.2 Effect of strain

As mentioned in the introduction section, experimental results with respect to the effect of strain on the In segregation are contradictory. To address this as well as to unravel the role of strain concerning the disagreement between experimental and our calculated results as discussed in the previous section, we apply bi-axial strain on the GaAs surface structures and determine In surface segregation energy using equation (1). Figure 5 displays the In surface segregation energies as a function of strain for various GaAs surface structures under consideration. For (110) and (100)$β2(2×4)$, In tendency to segregate towards the surface is reduced under applied tensile strain while it is enhanced under compressive strain. In the case of (111)$A$-Ga vacancy, a marginal decrease is observed for compressive strain whereas tensile strain has no effect on the In segregation. Further, the strain has a negligible effect for In segregation on (100)c(4×4) surface. From figure 5 it is clear that surface segregation energy of the β2(2×4) surface is lower than that of extreme As-rich c(4×4) structure at a compressive strain of 4% and is expected to decrease further with the increase in the compressive strain. Hence our results are in agreement with the earlier experimental observation of reduced In segregation for As-stabilized surface under applied strain conditions.

Above results indicate that strain plays a decisive role on In segregation for specific surface orientations, whereas it has insignificant effect for other orientations. This clarifies the conflicting experimental results on the effect of strain on segregation behavior. While Genet et al. [8] (also reference [15]) showed the negligible role of compressive strain on In segregation for As-stabilized GaAs(100) surface, our calculations confirm their results as the variation of segregation energy with respect to strain is negligible for extreme As-rich (100)c(4×4) surface reconstruction. Further, we may comment on other experiments which report significant enhancement of In segregation with increasing compressive strain [16–19], that their experiment used (100)β2(2×4) reconstruction, as this phase is found to be stable over a broad range of intermediate As and Ga-chemical potentials [25].

### 3.3 Bonding analysis

Chemical bonding analysis is performed to elucidate the In segregation behavior with respect to surface reconstruction and applied strain. We employed the concept of crystal orbital Hamilton population (COHP), which is a product of the density of states and the overlap Hamiltonian element [47]. The negative and positive values of COHP indicate bonding and anti-bonding, respectively. COHP calculations are performed using LOBSTER package [48,49], which extracts the bonding information from the results of DFT calculations. The COHP, integrated up to the Fermi level $E_f$ (ICOHP), is a measure of bond strength [50]. Calculated -ICOHP for Ga-As and In-As bonds in surface layer and in bulk for β2(2×4) and c(4×4) reconstructions as a function of applied strain are shown in figure 6. Predictably, -ICOHP of III-V bond decreases when Ga is substituted with In for all cases, implying decrease in the bond strength with In substitution.

Table 1 lists the -ICOHP values at zero strain conditions for Ga-As and In-As bonds in surface and bulk layer for β2(2×4) and c(4×4) reconstructions. The change in -ICOHP with In substitution -ΔICOHP (ΔICOHP= ICOHP(In-As)$_{n\text{-layer}}$ - ICOHP(Ga-As)$_{n\text{-layer}}$) is also listed. The -ΔICOHP is lower for surface than for bulk for both reconstructions indicating bond weakening due to In substitution is less pronounced for surface than for bulk. In other words, In prefers to segregate to surface than bulk, as expected. Further comparing -ΔICOHP values of c(4×4) and $\beta$2(2×4) reconstructions, we can say that the reduction in bond strength due to In substitution is lower for c(4×4) surface than for $\beta$2(2×4) surface. This gives the reason for the higher tendency of In surface segregation for c(4×4) than $\beta$2(2×4) reconstruction under zero strain conditions and thus explains the dependency of In surface segregation on the surface reconstruction.

Table 1: Calculated -ICOHP values for nearest-neighbor interactions Ga-As and In-As bonds in surface and bulk layer for $\beta$2(2×4) and c(4×4) reconstructions of (100) surface, at zero strain conditions.

| Type | -ICOHP (eV) | |
|---|---|---|
| | $\beta$2(2×4) | c(4×4) |
| **Surface:** | | |
| Ga-As | 3.97 | 4.60 |
| In-As | 3.41 | 4.20 |
| -ΔICOHP | -0.56 | -0.40 |
| **Bulk:** | | |
| Ga-As | 4.25 | 4.27 |
| In-As | 3.93 | 4.03 |
| -ΔICOHP | -0.32 | -0.24 |

Further to explain the effect of strain, variation of -$\delta$ICOHP for III-As bond (In-As bond in surface layer and Ga-As bond in bulk) as a function of applied strain, taking zero strain as reference ($\delta$ICOHP = ICOHP$_{x\%}$ - ICOHP$_{0\%}$) for both reconstructions are illustrated in figure 7. Interestingly, for $\beta$2(2×4) reconstruction the value of -$\delta$ICOHP for III-As bond in surface layer stays roughly constant with strain, indicating negligible influence of strain on In-As bond strength. Whereas in bulk layer, -$\delta$ICOHP increases with compressive strain and decreases with tensile strain. Since compressive strain strengthens and tensile strain weakens III-As bond in bulk, as a consequence surface segregation tendency enhances with compressive and reduces with tensile strain for $\beta$2(2×4) reconstruction. On the other hand, for c(4×4) reconstruction, -$\delta$ICOHP, and thus bond strength increases with compressive strain and decreases with tensile strain for III-As bonds in both bulk and surface layer. As a result, the change in surface segregation tendency with applied strain is marginal for c(4×4) reconstruction.

While many experimental measurements and theoretical models are available in the literature concerning In segregation in III-V alloys and interfaces, no quantitative results are presented for its strong dependency on surface reconstruction and strain. Such data is pivotal for the efficient designing of microelectronic and optoelectronic devices. In this regard, the results presented here may provide a guide to further experimental efforts to develop an efficient way to suppress the compositional broadening of the interface due to In segregation.

## 4. Summary

In summary, we have examined the In segregation on four stable surfaces of GaAs, using first-principles calculations based on DFT. The calculated surface energy of GaAs(110) is less than that of InAs(110), thus confirming In segregation on GaAs surface, which is in agreement with earlier experimental and theoretical reports. Our results show the strong dependence of In segregation energies on the surface orientation of GaAs. For (110) and (100)$\beta$2(2×4) structures, In tendency to segregate towards the surface is reduced under applied tensile strain while it is enhanced under compressive strain. The effect of strain is found to be negligible for (100)c(4×4) and (111)$A$-Ga vacancy reconstructions. These segregation energy trends are explained by considering III-V atomic bonding in surface and bulk layers. The results presented in this work could be valuable in experiments to achieve control on In segregation by managing As flux together with the application of strain.

**Figures**

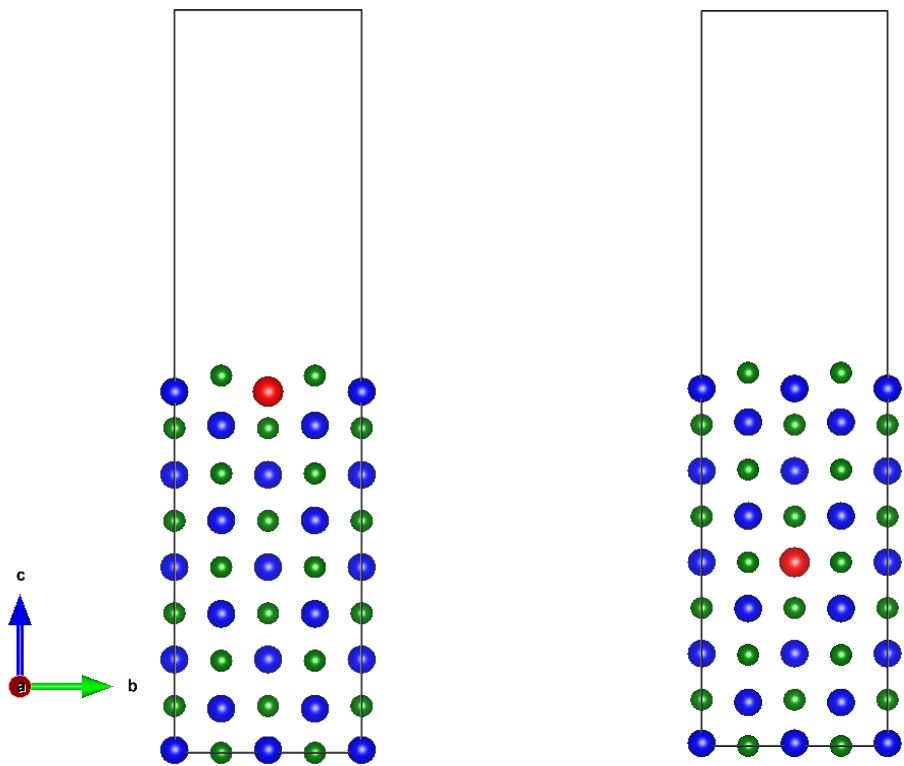

**Figure 1:** GaAs(110) slab model with one Ga atom replaced by In atom in (a) topmost layer and (b) 5$^{th}$ layer (bulk layer). Blue, red and green balls represent Ga, In and As atoms, respectively.

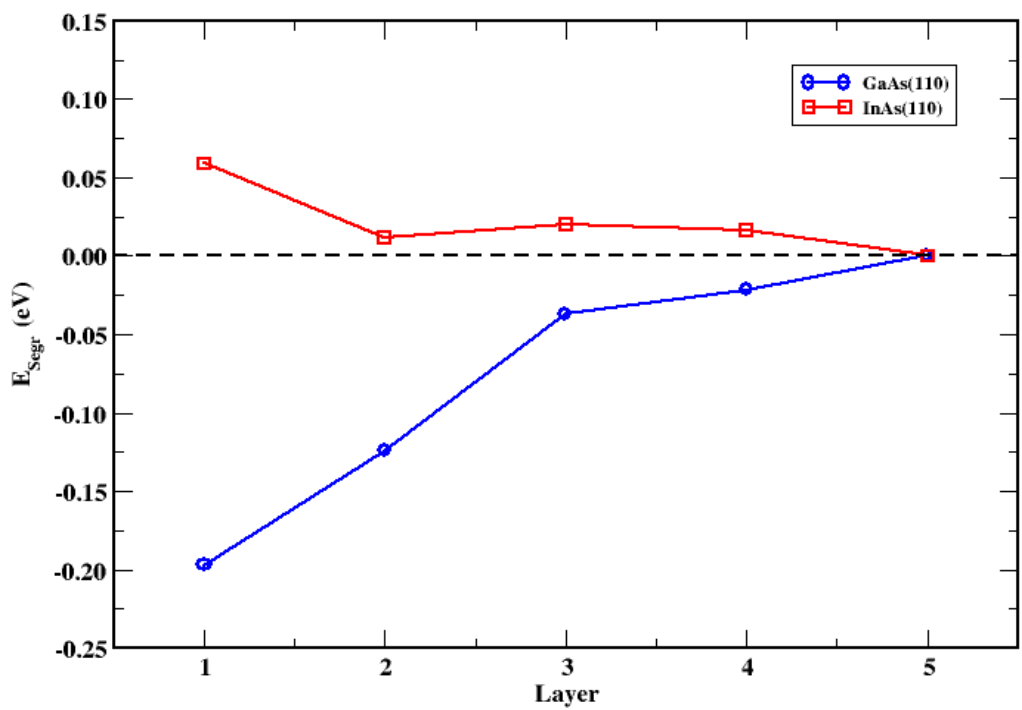

**Figure 2:** Segregation energy estimated using equation (1) for In in GaAs(110) and Ga in InAs(110) surfaces located at various atomic layers n (1-5).

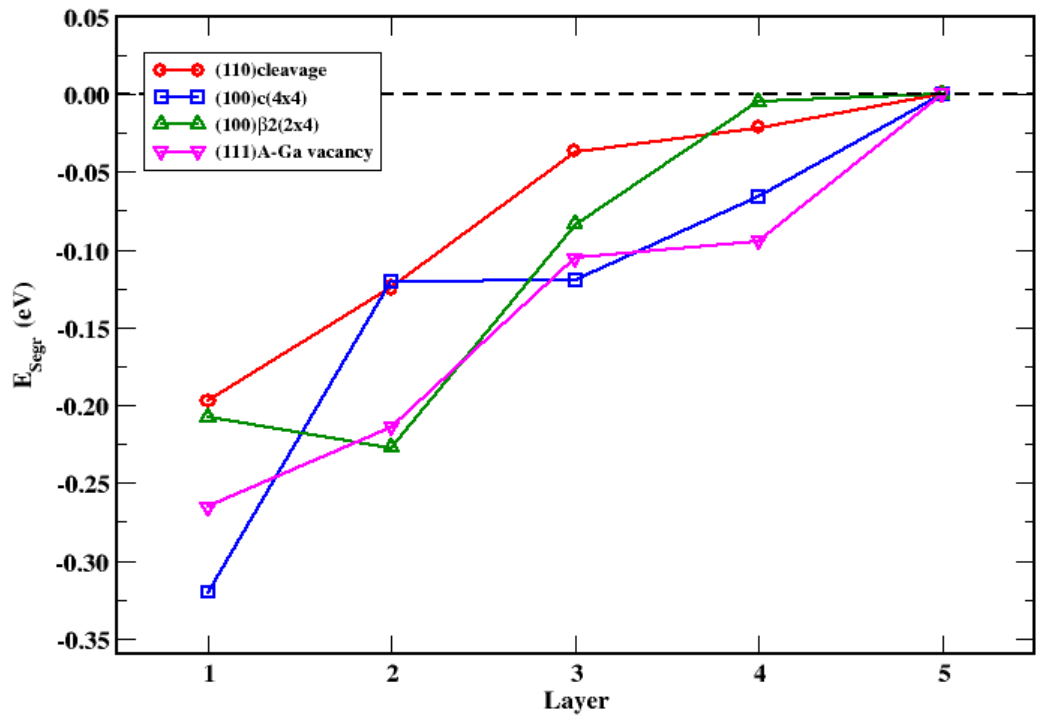

**Figure 3:** Segregation energies of In atom located at various atomic layers n for GaAs surfaces.

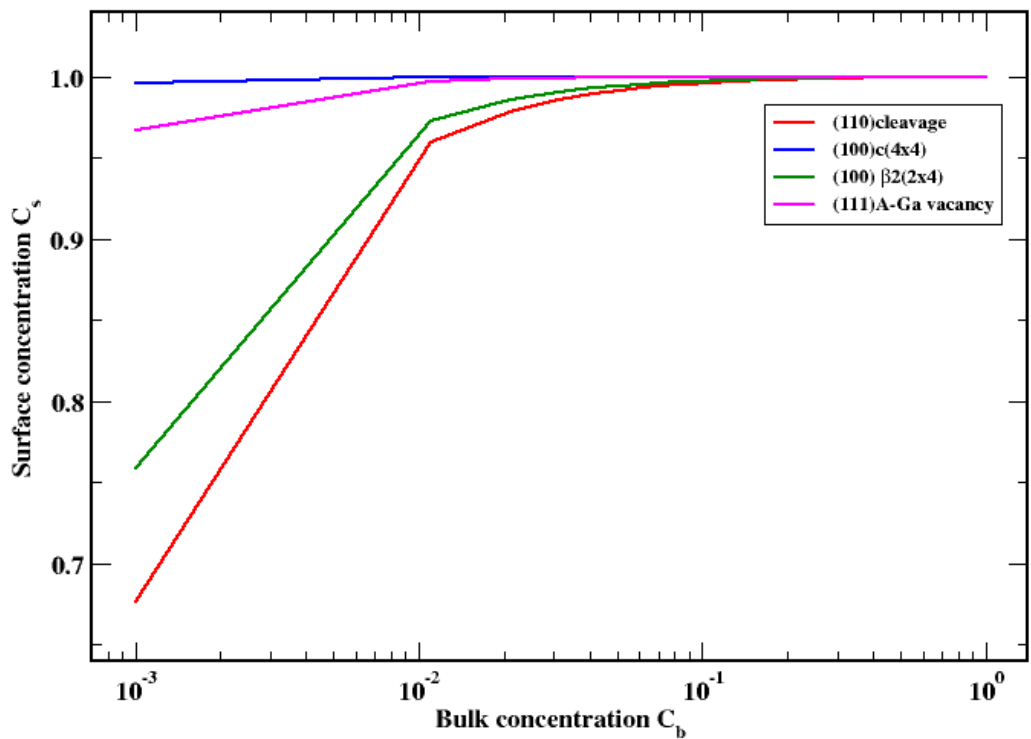

**Figure 4:** Surface concentration of indium as a function of its bulk concentration at 300 K for various GaAs surfaces.

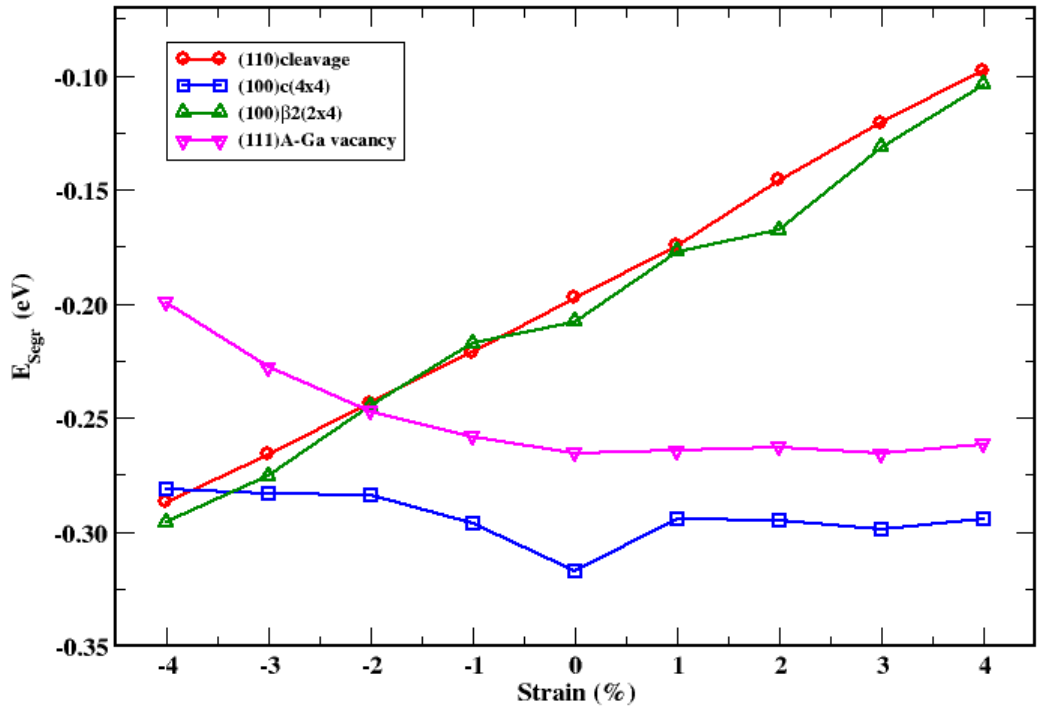

**Figure 5:** Surface segregation energies as a function of strain for various GaAs surfaces.

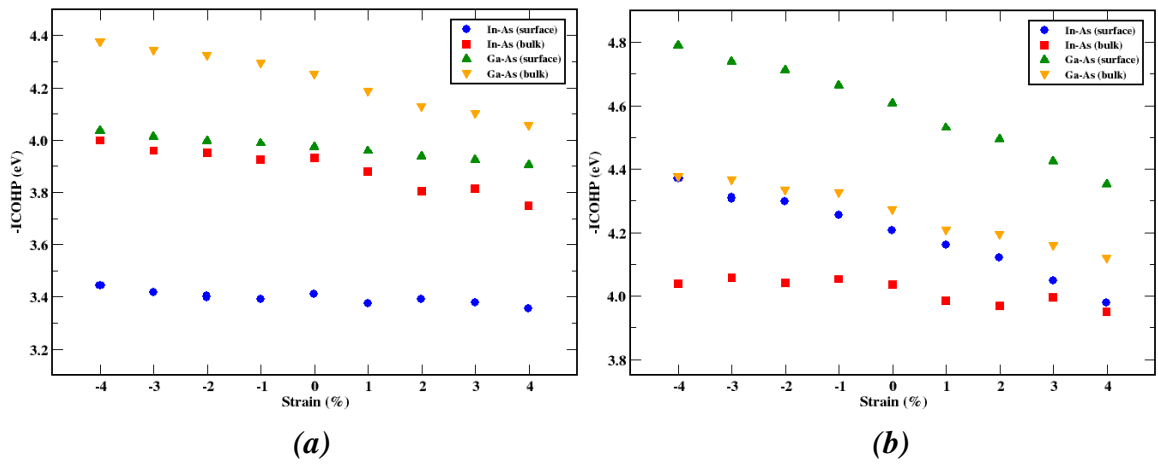

**Figure 6:** Calculated -ICOHP as a function applied strain for (a) (100)*β*2(2×4) and (b) (100)c(4×4) reconstructions of GaAs.

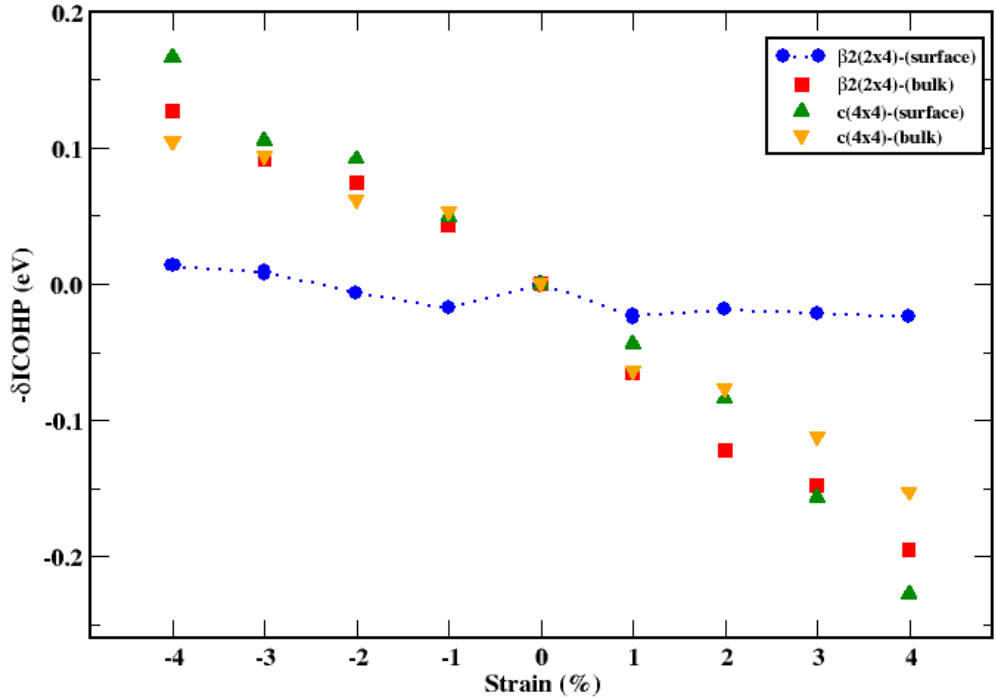

**Figure 7:** Variation of -δICOHP for III-As bond as a function of applied strain for (100)β2(2×4) and (100)c(4×4) surfaces.